\documentstyle[12pt]{article}
\input{psfig}

\title{ \large \bf Dynamics of Clusters in Two-dimensional Potts Model 
\thanks{This project is partially supported by Turkish 
Scientific and Technical Research Council (T\"{U}B\.{I}TAK) 
under the project TBAG-1141 and TBAG-1299.}}
\author{Y. G\"{u}nd\"{u}\c{c} and M. Ayd{\i}n \\
{\normalsize Hacettepe University, Physics Department, 
06532  Beytepe, Ankara, Turkey }}
\date{\normalsize\today}
\begin{document} 	
\begin{titlepage}

\titlepage

\maketitle

\vskip 0.5cm
\begin{abstract}
{\normalsize Dynamical behavior of the clusters during relaxation is
studied in two-dimensional Potts model using cluster algorithm.
Average cluster size and cluster formation velocity are calculated
on two different lattice sizes  
for different number of states during initial stages of the Monte
Carlo simulation. Dependence of these quantities on the order of the 
transition provides an efficient method to study  nature of 
the phase transitions occuring in similar models.}  
\end{abstract}

\end{titlepage}
\pagebreak

\section{Introduction}

\pagenumbering{arabic}

Recent interest in models possessing weak first-order phase
transition has increased the importance of distinguishing 
first- and second-order phase transitions for the simulation type of studies. 
One of the most successful methods to determine the order of the transition is
introduced by Challa et. al.~\cite{Challa:1986}. This method is based
on comparing the shape of the energy probability distribution of the
system with a single or double Gaussian for first- and second-order
transitions respectively.  Similarly, observing a double minima or
single minimum in free energy may lead to conclusive evidence on
the order of the transition ~\cite{Lee:1990}. However, for a finite lattice with
large correlation length, it is very difficult to determine  the order of
the phase transition even if one uses a series of lattices with 
increasing sizes.

$\;$

The information during a Monte Carlo run propagates with the
relaxation time which is proportional to a power of the correlation
length.  During first few iterations, this finite propagation speed
reveals physics of different time scales. This approach has
successfully used to investigate perturbative and non-perturbative
effects in 2-dimensional $O(3)$ sigma and $CP^{(N-1)}$
models~\cite{DiGiacomo:1992}, as well as $SU(2)$ and $SU(3)$ lattice
gauge theories~\cite{Alles:1993}.  The same idea is also applied to
$(2+1)$- dimensional scalar field theories to study the dynamics of
first-order phase transitions~\cite{Gleiser:1994}. In this work
similar ideas will be used to devise tools of distinguishing first-
and second-order phase transitions in spin systems by simply studying
the evolution of some observables in the first couple of hundred Monte
Carlo iterations. In this approach the major concern is the choice
of observables; since the global quantities are most sensitive
to the correlations in the system, their effects can be seen clearly
by use of such observables. Two of the possible candidates of
these observables may be the cluster size and the cluster formation
velocity. These quantities show different behaviour depending on the
magnitude of the correlation length. Large correlation length results
in faster initial thermalization, while in the case of short range
correlations the convergence is very slow; hence initial iterations in
a Monte Carlo simulation possess vital information on the phase
structure of the system.  The aim in this work is to extract
information about the phase transition by observing cluster (growth)
dynamics in two-dimensional $q$-state Potts model during early stages
of a Monte Carlo simulation.

$\;$

The Hamiltonian of the two-dimensional Potts model
\cite{Potts:1952} is given by
\begin{equation}
     {\cal H} = K \sum_{<i,j>} \delta_{\sigma_{i},\sigma_{j}}.
\end{equation}

Here $K=J/kT$ ; where $k$ and $T$ are the Boltzmann constant and 
the temperature respectively, and $J$ is the magnetic
interaction between spins $\sigma_{i}$ and $\sigma_{j}$, which can take
values $1,2, ..., q$ for the $q$-state Potts model.
In a Monte Carlo simulation  
the average cluster size (CS) can be calculated by taking
the average of $N_c$ clusters per iteration,  
\begin{equation}
CS = \displaystyle{{1}\over{N_c}}< \sum_{i=1}^{N_c} C_i >  
\end{equation}
where $C_{i}$ is the number of spins in the ${\rm i^{th}}$ cluster.
Being a function of the largest cluster,
the order parameter $(OP)$ is also  
a  global observable and it can be calculated through the relation   
\begin{equation}
OP=\frac{q\rho^{\alpha}-1}{q-1},  
\end{equation}
where $\rho^{\alpha}=N^{\alpha}/L^{D}$, $N^{\alpha}$ is the number of
spins in state $\sigma=\alpha$, $L$ and $D$ are the linear size and the
dimensionality of the system respectively.

$\;$

The two-dimensional, $q=2,3$ and $4$ state Potts model exhibits
second-order phase transition, while for $q \ge 5$ the transition is
known to be first-order \cite{Baxter:1973,Wu:1982}.  For $q=5$ the
correlation length is finite but very large and it decreases with
increasing $q$.  In this work the two-dimensional $q$-state Potts
model ($q=2,3, \dots , 8$) is chosen as the test ground for the above
mentioned ideas.  It is believed that comparisons of the 
behavior of the cluster growth or
the change in the order parameter  (per Monte Carlo
iteration)  for different $q$ can give some information 
about the degree of the phase transition.

$\;$

In the present work, the two-dimensional $q$-state Potts model
for $q=2$ to $8$ has been simulated during thermalization,
using a cluster update algorithm. Dynamical information about the
cluster formation is to be extracted through the calculation of
the cluster size. Commonly used quantities like energy,
order parameter and the relaxation time are also obtained.  The
organization of the article is as follows: Section 2 gives
general information about the cluster dynamics during
relaxation.  Results and discussions are given in section 3,
and the conclusions are presented in section 4.

\section{Dynamics of clusters in a spin system}

When a spin system is set to relax at a given temperature, spins tend
to form clusters with different sizes and shapes, depending on the
temperature and the strength of the interaction between the spins.
This formation occurs either by integration (forming large clusters)
or by disintegration (breaking into smaller clusters).  If the system
is ordered initially, first few Monte Carlo iterations immediately
break the configurations and small clusters start to appear. This
process gradually slows down as the system relaxes and reaches the
``equilibrium''. For a disordered start, the system remains as it is for the
first couple of iterations, while the information still has not been
propagated to the distances long enough to form clusters.  After a few
iterations, clusters at reasonable sizes start to appear and their
average size increases as the system relaxes towards equilibrium.

$\;$

To check how the thermodynamic quantities approach the equilibrium
values or more precisely how the cluster size changes with increasing
number of iterations, one can start Monte Carlo runs from different
starting configurations.  The average of a quantity like the cluster
size over all different starting configurations at every iteration
gives the time dependent ensemble average of that quantity. If the
correlation length is very large, sizable clusters immediately grow
and the cluster size, after initial few tens of iterations, fluctuates
around an average value.  On the contrary, for the systems with
correlation length smaller than the lattice size, this thermalization
requires very long Monte Carlo runs and a distinguishing aspect is
that fluctuations around the iteration-dependent average value are
very large. The formed large clusters can not maintain their sizes
around an average value for many iterations; they break into smaller
clusters contributing the fluctuations in the
system~\cite{Aydin:1996}.  For a given $q$, average cluster size
reaches a different equilibrium value for different temperatures and
the number of Monte Carlo iterations to reach the equilibrium depends
on the temperature at which the simulation is performed.  To avoid
ambiguities, all simulations in this work are performed at 
the temperature $ T = T_c $ where the specific heat peaks are observed.

$\;$

A possible function for the relaxation process may be
~\cite{Gleiser:1994}
\begin{equation}
f(t)=f_{0} [1 - \exp {(-t / \tau)^{\theta}}]
\label{f}
\end{equation}
where $f_{0}$ is the final value of the function, $\tau$ is
the relaxation time which is related to the inverse of the
correlation length,  $\theta$ is some exponent  
and $t$ indicates the number of iterations (Monte Carlo time).   
By fitting
the time dependent average cluster size data to equation (4),
one can obtain  quantities $f_{0}$, $\tau$ and $\theta$. 
The change in the
average cluster size per iteration,  which may be called  the
``cluster formation velocity''$(CFV)$, gives another important information of
the same type. This quantity can be calculated
through the relation
\begin{equation}
 CFV = {{\partial f(t)}\over{\partial t}} 
\end{equation}
which is simply the derivative of the  function $f(t)$ which represents 
the mean cluster size and is given in  equation (4), 
or the derivative of the fitted polynomial function,
with respect to Monte Carlo time.   
Cluster formation velocity gives a clear  
indication of how fast the clusters are formed and how many
iterations can be used for the initial thermalization of
the system.  Hence it is related to the dynamics of the spins
and to the correlation length in the system.

\section{Results and discussions}

Averages of energy, order parameter and cluster size during relaxation
of the system are obtained in the two-dimensional Potts model with
number of states $q$ varying from $2$ to $8$, using the cluster flip
algorithm which was first introduced by Swendsen and
Wang~\cite{Swendsen:1987} and later modified by
Wolff~\cite{Wolff:1989}.  The algorithm used in this work is the same
as Wolff's algorithm with the exception that before calculating the
observables, searching the clusters is continued until the total
number of sites in all searched clusters is equal to or exceeds the
total number of sites in the lattice.  The averages are calculated
about $5000$ replicas of the system starting from different disordered
initial configurations and errors are calculated using jackknife error
analysis.  All the computations are performed at the finite size
critical value $K_c$ of the coupling $K$ on $32 \times 32$ and $64
\times 64$ lattices. $K_c$ values are obtained as the values
corresponding to the maxima of specific heat calculated at a previous
work, where the cluster size related measurables are studied at
equilibrium~\cite{Aydin:1996}.

$\;$

Figure 1 shows plots of average cluster size ($CS$) calculated
dynamically during the thermalization for $q=2$ to $8$ on $64 \times
64$ lattice. Results for the $32 \times 32$ lattice give basically the
same information with higher slopes for early iterations or with
faster thermalizations.  These plots show that the slope of the curves
at early iterations decreases as $q$ increases, which is consistent
with the discussion given in Section 2.  Large slopes in $CS$ data for
$q<5$ (i.e. when the transition is second-order) indicate fast
formation of the clusters. For $q \ge 5$, phase transition is
first-order and the correlation length decreases as $q$ increases,
therefore the cluster formation is rather slow.

$\;$

Dynamical data for energy, order parameter and average cluster size
calculated on small and large lattices with different $q$ are fitted
to both exponential (equation(4)) and the polynomial functions.  These
functions are used to calculate cluster formation velocity ($CFV$).
$CFV$ shows variation in $CS$ as a function of the number of
iterations, so it is simply calculated by taking derivative of the
fitted function (equation(5)) with respect to Monte Carlo time.
Figure 2 shows $CFV$ for the large lattice for $q$ values varying from
$2$ to $8$. For small $q$, clusters form very quickly, so maximum of
$CFV$ occurs at a higher value than that of large $q$. As these plots
show, $CFV$ has a distinct maximum for $q < 5$ and it is in the form
of a curve with a broad maximum for $q=5$ and $6$.  It is observed
that starting from $q=7$, it is difficult to find a maximum for $CFV$
data.  For $q=8$ the maximum is at about the first or second
iteration, and after a few iterations, $CFV$ values are almost
constant.  The runs performed for $q=9$ and $10$ show the same
behaviour.  As it is seen from figure 2, number of iterations where
the maximum of $CFV$ occurs increases as $q$ increases.  This
information is consistent with the observation that cluster formation
is slow for large $q$.  $CFV$ plots for the $32 \times 32$ lattice
show similar variations, the only difference being that the maximum
points of $CFV$ occur at higher values than that for the large lattice
for each $q$ value.

$\;$

Relaxation time $\tau$ and the exponent $\theta$ for each $q$ are
calculated through exponential fit by use of equation(4).  $\theta$
values are related to initial behaviour of the relaxation process. It
is observed (both for $32 \times 32$ and $64 \times 64$ lattices) that
$\theta$ is around a value of about $2.0$ for $q=2$ and decreases
monotonically as $q$ increases, reaching a value of about $1.0$ for
$q=7$ (and lower for $q=8$).  To see the variation of $\tau$ for
different $q$ on small and the large lattice, its ratio to $\tau$ for
$q=2$ is plotted on a logarithmic scale. Plots of $\tau(q=2)/\tau(q)$
versus $q$, for small and large lattices are shown in figure 3.  One
can see from this figure that the slope of $\tau(q=2)/\tau(q)$ remains
almost constant for the smaller lattice, while the slope of the ratio
calculated on the larger lattice continuously changes with increasing
$q$ for $q\ge5$, as an indication of the sensitivity of $\tau$ to
decreasing correlation length. In other words, the $\tau$ values for
different size lattices are parallel within error limits for small $q$
($q=2, 3$ and $4$), and starting from $q=5$, the discrepancy between
two curves increases as $q$ increases.  $\tau$ values calculated using
the order parameter ($OP$) data is also shown in figure 3.  As it is
clear from figure 3, their behaviour as a function of $q$ is very
similar.  For comparison, $CFV^{max}(q)/CFV^{max}(q=2)$ versus $q$ for
small and the large lattice is plotted on a logarithmic scale as well
(figure 4), and it gives similar information about the $q$ dependence
of the cluster size and the order parameter data given in figure 3.
Since the definition of the order parameter involves all
spins in the same orientation (throughout the lattice rather than
individual clusters) the order parameter can not be used to calculate
a quantity similar to  $CFV$. Hence in figure 4 the information
comes from the cluster size data only.

$\;$

If the phase transition is second-order (i.e. the correlation length
is infinite), on a finite lattice, the lattice size sets the limits of
the correlation length. Hence for $q=2, 3 $ and $4$ $\tau$ ratios for
different sizes exhibit this limit and this ratio is about $2.0$. If
the correlation length is larger than the lattice size, the
measurements of any quantity which is proportional to correlation
length will lead to a value proportional to the lattice size.  For $q
\ge 5$, since the correlation length is finite, the discrepancy starts
to appear.  The behaviour of a local operator such as energy is very
different and one can not observe the above discussed changes as $q$
varies.

\section{Conclusions}

Dynamical behaviour of clusters is studied in two-dimensional Potts
 model for the number of states $q$ varying from $2$ to $8$ on $32
 \times 32$ and $64 \times 64$ lattices.  Dynamical data for energy,
 order parameter and the cluster size ($CS$) are obtained during relaxation
 of the system at the finite size critical coupling value $K_c$ and
 this data is fitted to both exponential and the polynomial functions
 for each value of $q$. Relaxation time $\tau$ (from the exponential
 fit) and the cluster formation velocity $CFV$ (from the exponential
 and polynomial fits) are obtained for small and large lattices.
$CFV$ curves given in figure 2 have distinct maxima for $q=2, 3$ and
$4$, while they have a broad maximum for larger $q$ (and almost
constant for $q=7$ and $8$).  This is consistent with the information
given by dynamical data shown in figure 1.  It is also observed in
figure 1 that the slope of the dynamical data decreases as $q$
increases. This observation, together with the information from the
exponential fit ($\tau$ versus $q$), enhances the idea that the
clusters form quite slowly for large $q$ starting from $q=5$, for
which case the transition is first-order.

On a finite lattice, if the transition is
second-order ($q=2,3,4$), the correlation length is infinite and the
size of the system sets the limit on the correlation length. 
This is consistent with the information obtained 
when $\tau$ and $CFV$
measurements on different sizes are compared.  
It is  observed that their ratios for different sizes are about to be the same
for small $q$ ($q=2,3,4$); close to the ratio of the linear sizes when
calculated directly.  
For $q \ge 5$, although the  
correlation length is larger than both lattice sizes   
(especially for $q=5$ and $6$), 
the size ratios start to deviate from $2.0$ significantly,
corresponding to the discrepancy between two curves in
figures 3 and 4.  
This behaviour is indicative of the sensitivity of the measured quantities
to changes in the correlation length  
(hence to  changes in the order of the transition). 
This means that  
cluster size related measurables like
$CFV$ can give some relevant information about the nature of the
transition in the Potts model  and in similar models 
(using only a first few hundreds of the
Monte Carlo iteration).

$\;$

It should be stressed here that the $\tau$ values calculated from
autocorrelations obtained as a result of the runs of $1-2
\times 10^6$ iterations (after thermalization) at $K_c$ are the
same as the ones obtained from dynamical data, within the error
limits.  This shows that the results of cluster formation
dynamics are reliable, and the quantities like $CS$, $\tau$
and $CFV$ obtained during early iterations can give reliable
information about the critical behaviour of the system.  When
computing times to calculate autocorrelations and the dynamical
data are compared, it is easy to see that critical behaviour of
similar systems can be studied using this new method, and the
computational effort is much less than that for the standard
and commonly used methods.  

\vskip 1cm

\section*{Acknowledgements}

Authors thank T. $\c{C}$elik  for reading the manuscript. 

\pagebreak

\pagebreak

\centerline{FIGURE CAPTIONS}

\begin{description}

\item {Figure 1.} The average cluster size calculated dynamically ($CS$)
  for $q=2,3, ..8$ on $64 \times 64$ lattice.      

\item {Figure 2.} Cluster formation velocity ($CFV$)  for $q=2,3, ...8$
 calculated on  $64 \times 64$ lattice.   

\item {Figure 3.} $\tau(q=2)/\tau(q)$ versus $q$ ($q=2,3, ...8$)   
  for cluster size ($CS$) and the order parameter ($OP$)
  calculated on $32 \times 32$ and $64 \times 64$ lattices.    

\item {Figure 4.}  $CFV^{max}(q)/CFV^{max}(q=2)$  
   versus $q$ ($q=2,3, ...8$)   for cluster formation velocity ($CFV$)  
       calculated on $32 \times 32$ and $64 \times 64$ lattices. 

\end{description}

\pagebreak

\begin{figure}
\psfig{figure=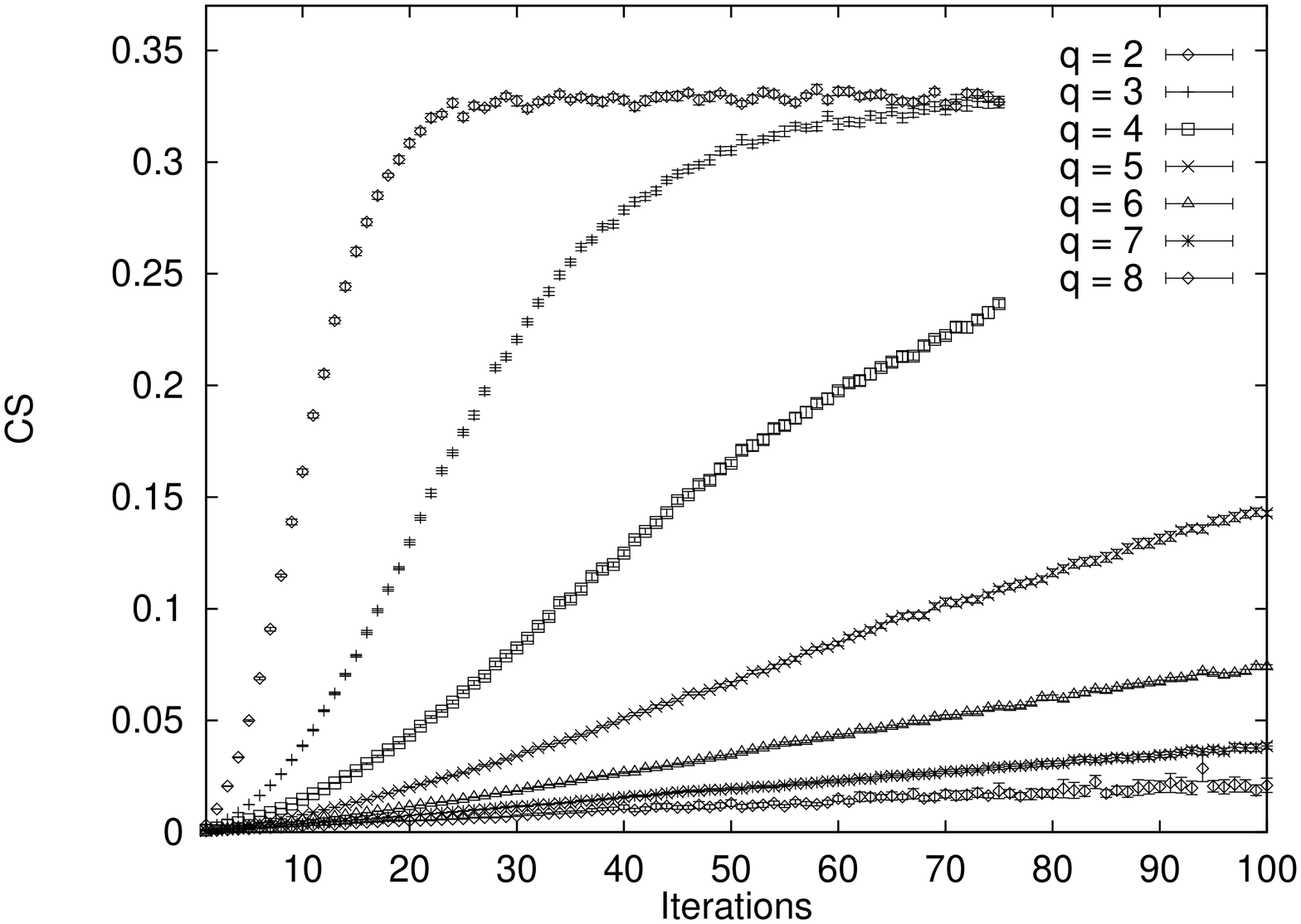,height=8cm,width=12cm,angle=-90}
\caption{}
\end{figure}
\begin{figure}
\psfig{figure=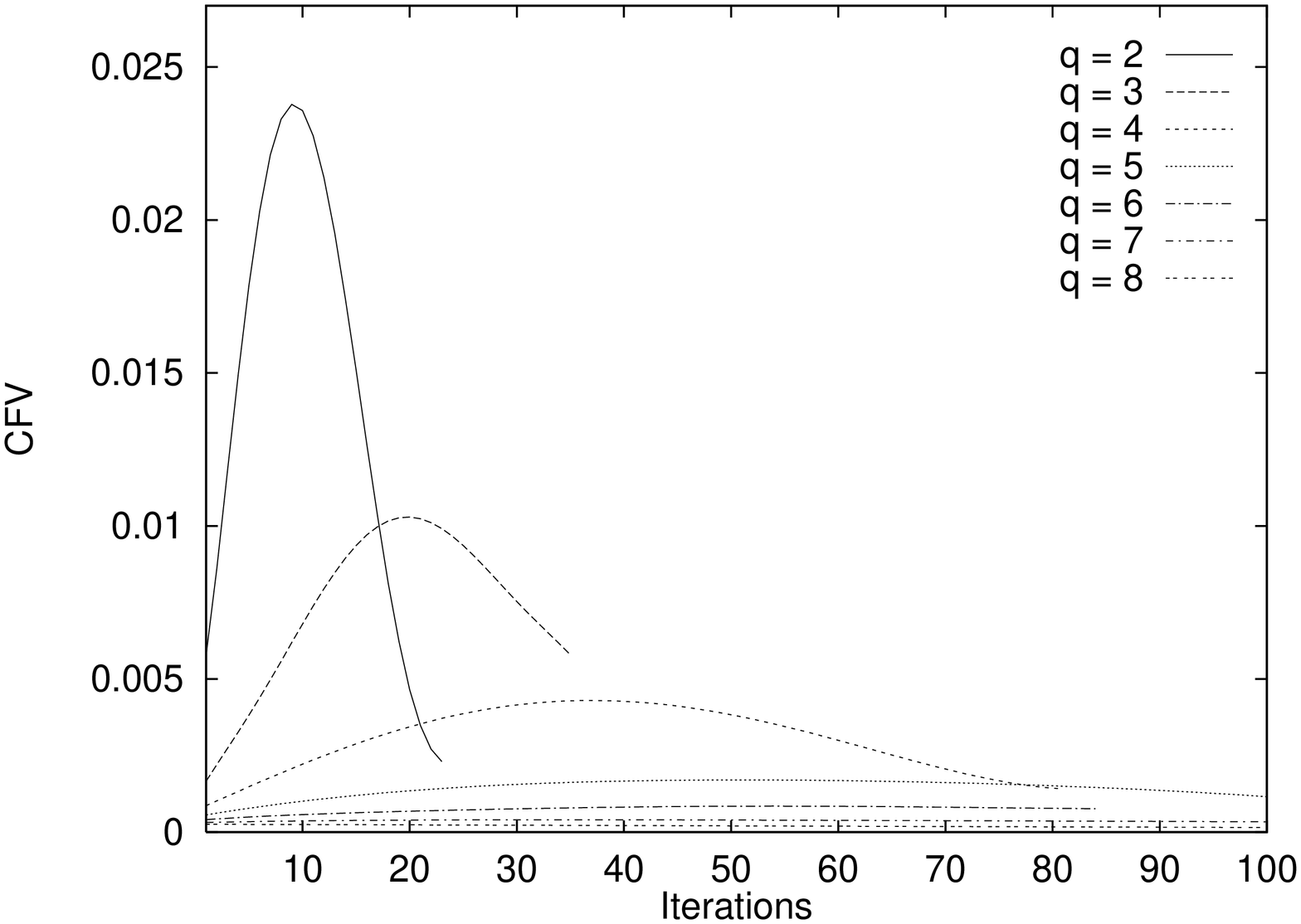,height=8cm,width=12cm,angle=-90}
\caption{}
\end{figure}
\begin{figure}
\psfig{figure=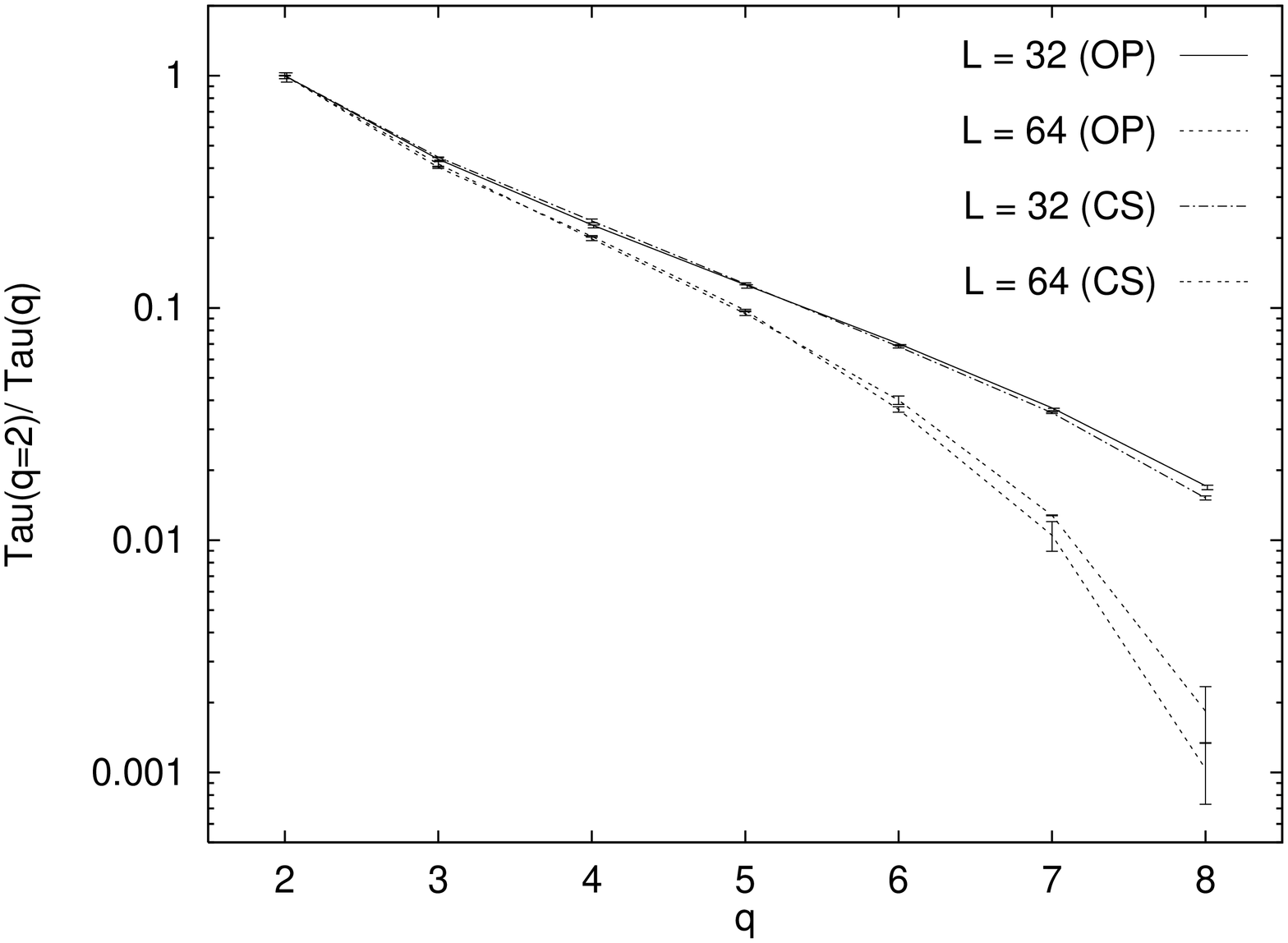,height=8cm,width=12cm,angle=-90}
\caption{}
\end{figure}
\begin{figure}
\psfig{figure=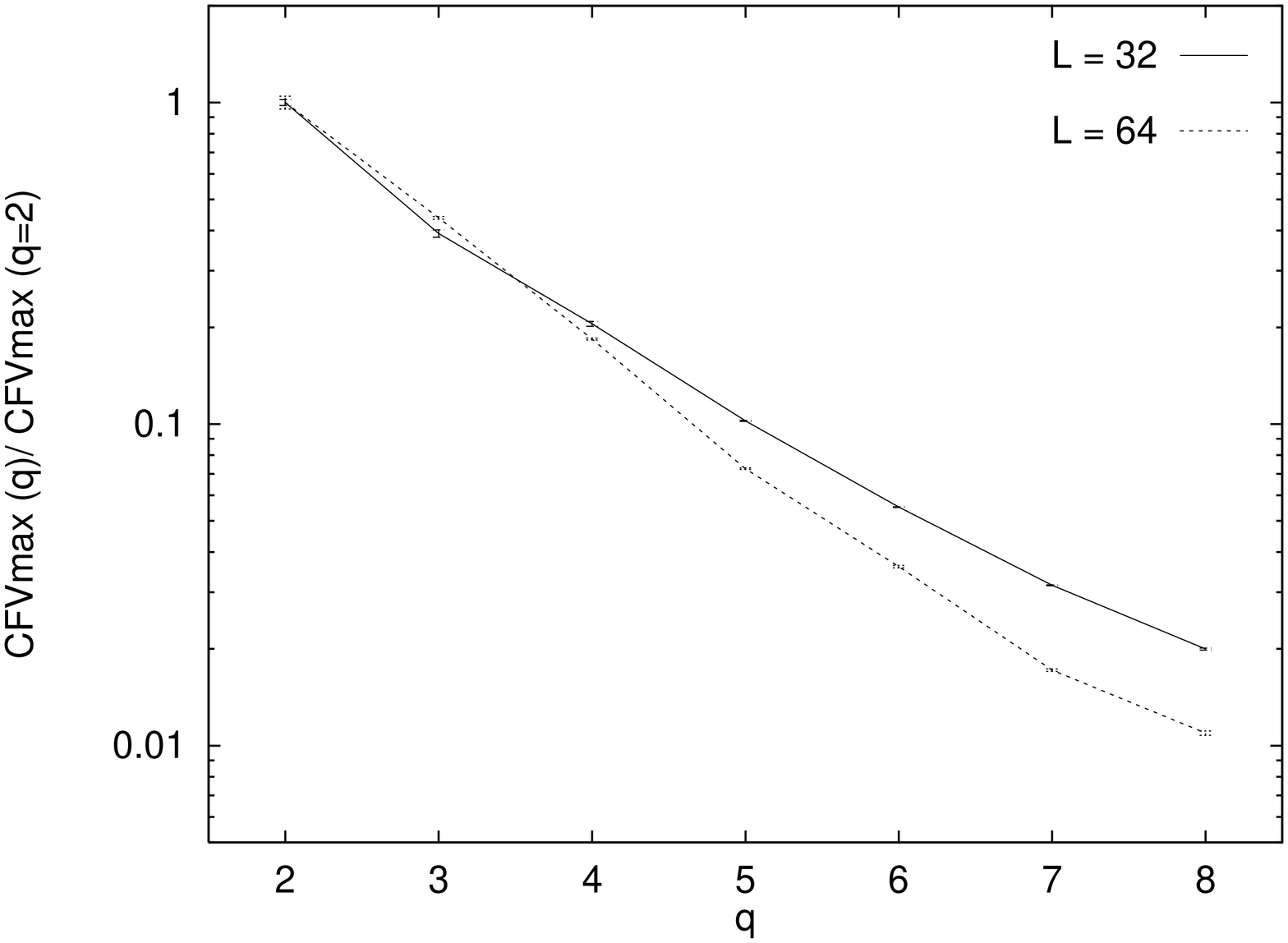,height=8cm,width=12cm,angle=-90}
\caption{}
\end{figure}
\end{document}